\documentclass[11pt,a4paper]{article}

\usepackage[utf8]{inputenc}
\usepackage[T1]{fontenc}
\usepackage{amsmath,amssymb}
\usepackage{graphicx}
\usepackage{booktabs}
\usepackage{hyperref}
\usepackage[margin=2.5cm]{geometry}
\usepackage{xcolor}
\usepackage{natbib}
\usepackage{enumitem}

\title{Bitcoin Price Prediction: Peer-Reviewed Evidence\\
  and Social Media Discourse}

\author{
  Carlos Baquero\\
  \textit{INESC TEC \& Faculty of Engineering, University of Porto}\\
  \texttt{cbm@fe.up.pt}
}

\date{}

\begin{document}
\maketitle

\begin{abstract}
Bitcoin price prediction has attracted hundreds of academic papers and
continuous social media debate, yet the field lacks consensus on even
basic questions: can any model beat a naive ``today's price'' baseline
at horizons of one to six months?  We survey the peer-reviewed
landscape, categorize papers by evaluation methodology, and contrast
academic findings with informal but substantive discourse on
X/Twitter.  The picture that emerges is sobering.  At short-to-medium
horizons, no peer-reviewed study has shown robust superiority over
the naive baseline across multiple market regimes.  Daily
predictability is real but does not extend to hourly or monthly
horizons, and may not survive transaction costs.  The stock-to-flow
model has failed formal out-of-sample testing, and Metcalfe's Law
valuations have been challenged as spurious.
The Bitcoin price power law, while empirically compelling, has not
been subjected to formal distributional tests.  Meanwhile, social
media practitioners raise valid statistical critiques---ordinary
least squares (OLS) violations, backtest overfitting, spurious
regressions---that the academic literature has not formalized.  We
identify open research directions and propose concrete methodological
standards for future work---walk-forward evaluation, multi-regime
holdout windows, naive baseline comparison, inclusion of zero in
hyperparameter grids, and Diebold-Mariano significance
testing---arguing that the field's primary need is not more models
but better evaluation.
\end{abstract}

\section{Introduction}

Bitcoin's price trajectory has attracted modelling efforts from
physics, computer science, economics, and machine learning.  A Google
Scholar search for ``Bitcoin price prediction'' returns thousands of
results, growing by hundreds per year.  Yet the predictive value of
this body of work is unclear: most papers evaluate on a single
chronological train/test split~\citep{mcnally2018predicting}, use
in-sample metrics only~\citep{katsiampa2017volatility}, or fail to
compare against the simplest possible baseline---today's
price~\citep{kose2025deep}.

This evaluation gap matters because Bitcoin is a non-stationary asset
whose market structure has evolved substantially across its short
history~\citep{lo2004adaptive}.  A model that works in one market
regime (e.g., the 2017 bull run) may fail catastrophically in another
(e.g., the 2018--2019 crypto winter).  Single-split evaluations
cannot detect this failure mode.

A parallel discourse exists on social media, particularly X/Twitter,
where quantitative practitioners and independent researchers---some
with academic credentials---raise methodological critiques of Bitcoin
models.  These critiques---concerning ordinary least squares (OLS)
regression assumptions, backtest overfitting, and spurious
regressions---are often technically valid but have not been
formalized in peer-reviewed venues.

This paper surveys the landscape with four goals: (1)~distinguish
validated results from noise in the peer-reviewed literature,
(2)~document the gap between academic publications and social media
discourse, (3)~identify open research directions, and (4)~propose
concrete methodological standards for future work.

\section{Survey methodology}

We identified key papers through targeted searches on Google Scholar
and Web of Science for ``Bitcoin price prediction/forecasting''
(2017--2026), supplemented by forward and backward citation tracing
and domain expertise.  Selection combined two inclusion criteria:
papers with out-of-sample evaluation (selected positively for
methodological rigor), and widely cited papers with weaker
methodology (included to illustrate the field's state).  We
categorize papers into three evaluation tiers:

\begin{description}[leftmargin=2.5em]
  \item[Tier~1.] Walk-forward or rolling-window out-of-sample
    evaluation; the strongest studies in this tier also use a naive
    or random-walk baseline, though some rely on alternative
    benchmarks (see Appendix~\ref{app:studies}).
  \item[Tier~2.] Single chronological train/test split with
    out-of-sample evaluation.
  \item[Tier~3.] In-sample only, or model selection by information
    criteria (Akaike Information Criterion, AIC; Bayesian Information
    Criterion, BIC) on the full dataset.
\end{description}

For social media discourse, we conducted targeted searches on X/Twitter
using Grok (X's AI assistant with full post search) for discussions
of Bitcoin prediction methodology, prioritizing accounts with
verifiable credentials (PhD holders, university affiliations,
quantitative analysts) and posts referencing academic papers.
Social media posts are cited as footnotes with URLs and access
dates, not as peer-reviewed sources.  We note that Grok's retrieval
coverage is opaque and search reproducibility is limited; the
social-media component of this survey is therefore illustrative
rather than exhaustive.

We identified 23 papers meriting detailed discussion.  The full
Bitcoin prediction literature numbers in the hundreds, but the vast
majority fall into Tier~3 and contribute limited knowledge about
genuine predictive ability.

\section{The peer-reviewed landscape}

The Bitcoin prediction literature addresses several distinct tasks,
each with its own appropriate baseline and evaluation criteria.
Table~\ref{tab:tasks} summarises the taxonomy; the subsections that
follow are organised accordingly.

\begin{table}[t]
\centering
\small
\begin{tabular}{p{3.2cm}p{3.5cm}p{3.2cm}p{2.8cm}}
\toprule
Task & Naive baseline & Primary metric & Sections \\
\midrule
Price-level forecasting & Today's price & RMSE, MAE & 3.1, 3.2 \\
Return forecasting & Zero return & $R^2_{\text{OOS}}$, MSFE & 3.1, 3.2 \\
Direction classification & Random-walk sign & Accuracy, DA & 3.2 \\
Structural valuation & None (explanatory) & In-sample $R^2$ & 3.3 \\
Bubble / regime detection & No-bubble null & Precision, recall & 3.3 \\
\bottomrule
\end{tabular}
\caption{Taxonomy of Bitcoin prediction tasks.  Results from
  different tasks are not directly comparable; a model that
  ``beats the naive baseline'' for returns may fail for price
  levels, and vice versa.  Section~3.4 (the evaluation methodology
  crisis) cuts across all tasks and is not specific to any single
  row.}
\label{tab:tasks}
\end{table}

\subsection{Naive dominance at short-to-medium horizons}

Beyond very short-term microstructure effects (discussed in
Section~3.2), the strongest finding in the peer-reviewed literature
is a null result: no peer-reviewed study has shown a model that
reliably beats the naive baseline across multiple market regimes at
horizons of one to six months.  Throughout this subsection,
``naive'' means the task-appropriate benchmark from
Table~\ref{tab:tasks}: today's price for price-level forecasting,
zero return for return forecasting, and the random-walk sign for
direction classification.

\citet{puoti2024quantifying} conduct walk-forward backtesting of 12
models---ARIMA, Prophet, Random Forest, XGBoost, LSTM, NBEATS, and
others---on cryptocurrency prices at 1-, 7-, and 30-day horizons.
Naive models (NaiveDrift, NaiveSeasonal) consistently outperform all
complex alternatives.  They conclude that univariate cryptocurrency
forecasting is ``essentially comparable to pure noise forecasting.''

At monthly horizons, the evidence base narrows to a single preprint:
\citet{baquero2026naive}, currently under peer review, extends the
naive-dominance finding to 1--6 month horizons using five
non-overlapping holdout windows spanning 2016--2026.  Across 40
machine-discovered models and 18 Bayesian Structural Time
Series~\citep{scott2014predicting} configurations, training-data
optimisation selects zero correction in every holdout window---no
model beats the naive baseline across multiple market regimes.
Independent replication at monthly horizons is an open need
(Section~\ref{sec:open}).

\citet{arain2025forecast} address whether statistically significant
forecast improvements translate to economic value.  Using real-time
out-of-sample forecasts from daily bivariate VARs (October
2021--February 2024), they find that most predictors fail to generate
trading profits even when they show statistical significance.  Only
two of the 12 predictive indices they examine---the USD index and
the Shanghai Stock Exchange composite---yield significant excess
profits when used as Bitcoin trading signals, and only during
periods of large price swings.  Statistical predictability, they conclude, does not
generally imply economic value.

\citet{yae2022outofsample} test a wide range of in-sample predictors
(investor attention, trading volume, network metrics) for Bitcoin,
Ethereum, and Ripple returns.  ``Well-known in-sample predictors''
fail out-of-sample; only stochastic correlation with stock markets
yields modest out-of-sample $R^2$ (up to 2.7\% for Bitcoin).

\citet{cakici2024machine} confirm the complexity penalty in a
cross-sectional setting: across more than 500 cryptocurrencies,
OLS outperforms regression trees and neural networks for return
prediction out-of-sample---the benefits of model complexity are
limited even when the feature space is rich.

Even regime-aware models fail: \citet{agakishiev2025regime} combine
Hidden Markov Models with reinforcement learning, allowing model
parameters to vary across market regimes, yet detect no out-of-sample
improvement over simpler approaches.

These findings echo the Meese-Rogoff
puzzle~\citep{meese1983empirical}---the proposition that exchange
rate models cannot outperform the random walk
out-of-sample---transposed to cryptocurrency markets.  The picture
is not uniform: \citet{magner2022cryptocurrency} report that
autoregressive models augmented with lagged Bitcoin returns do beat
the random walk for some of the 13 major cryptocurrencies they
study, and note that cryptocurrencies are more persistent than
conventional exchange rates---a feature that, if anything, makes
the naive baseline harder to beat for Bitcoin itself.

\subsection{Short-term microstructure predictability}

At daily and sub-daily frequencies, some models do beat the random
walk.  This does not contradict the naive result at longer horizons;
rather, it reflects frequency-specific microstructure effects.

\citet{gradojevic2023forecasting} apply Random Forest to technical
indicators at hourly and daily frequency, using expanding-window
out-of-sample evaluation.  At hourly frequency, no model
significantly outperforms the random walk---weak-form
efficiency~\citep{fama1970efficient} holds.  At daily frequency, RF
with technical indicators beats the random walk significantly.  This
frequency boundary within ``short-term'' is itself informative: even
microstructure predictability has limits.
\citet{denicola2021intraday} documents a related phenomenon at
medium frequencies (1--4 hours): significant negative first-order
autocorrelation indicating systematic mean reversion, attributed to
investor overreaction and cascading liquidations.  This is an
in-sample characterisation rather than an out-of-sample forecasting
result (Tier~3), so it speaks to the existence of the mechanism
rather than to its tradability.

\citet{berger2024forecasting} compare ARMA-GARCH with RNN and LSTM
for daily Bitcoin returns using rolling-window out-of-sample
evaluation.  A simple RNN matches or outperforms ARMA-GARCH, but
LSTM does not improve over simple RNN---complexity does not help
even at daily frequency.

A distinct framing of the daily-frequency task is direction
classification.  \citet{kim2025direction} use CNN-LSTM to predict
the sign of daily Bitcoin price changes, incorporating stock-market
indices, commodities, and a maximum-drawdown indicator as features;
they then evaluate the resulting trading strategy by comparing its
maximum drawdown against an S\&P~500 buy-and-hold benchmark.  This
is a Tier~2 study (single chronological split) and reports no
naive-baseline comparison.

\citet{wei2023forecasting} test neural networks, SVM, and gradient
boosting with sentiment and volatility features, using rolling
six-month windows over 2014--2019.  Gradient boosting (XGBoost)
performs best, with formal significance confirmed via Diebold-Mariano
and Model Confidence Set tests.  However, the benchmark is the best
of 295 individual linear models, not a naive/random walk baseline;
the evaluation period is limited (pre-2020); and the 295-model pool
combined with ML selection creates a backtest overfitting risk per
\citet{bailey2014pseudo}.

\citet{gurgul2025nlp} provide the strongest case for NLP-augmented
forecasting: using BART zero-shot classification on news and social
media to detect bullish/bearish sentiment, they show consistent
improvements in profitability and Sharpe ratio across rolling-window
cross-validation folds at daily frequency, with BTC covered from
August 2011 and ETH from August 2015, both through March 2023.  The
result is notable, though the baseline is buy-and-hold rather than a
naive price forecast, results are not disaggregated by market
regime, and the evaluation ends before the 2023--2026 period.

A distinct source of short-term predictability is cross-cryptocurrency
spillovers.  \citet{guo2024cross} find that lagged returns of other
cryptocurrencies predict focal cryptocurrency returns out-of-sample,
with a long-short portfolio generating sizable returns after
transaction costs.  The mechanism---slow information diffusion due to
limited investor attention across a fragmented market---is consistent
with microstructure effects rather than fundamental predictability.

The synthesis is important: microstructure effects (bid-ask bounce,
delayed price discovery, order flow imbalances, cross-asset
spillovers) create exploitable autocorrelation at short horizons.
Whether this autocorrelation survives aggregation to monthly
returns is a separate question, and the present survey finds no
peer-reviewed evidence that it does.  \textbf{Daily predictability
does not imply monthly predictability.}

\subsection{Structural and valuation models}

A separate literature attempts to explain Bitcoin's price level
through structural models.  The most popular ones have failed formal
out-of-sample testing; the more theoretically motivated proposals
remain unreviewed.

\paragraph{Stock-to-flow.}
\citet{shelton2024bitcoin} shows that the stock-to-flow (S2F) model
explains Bitcoin returns in-sample but has ``limited to no ability''
to predict out-of-sample.  The model's significance vanishes once
time fixed-effects are introduced, revealing that its explanatory
power is confounded with the log-time trend (80.57\% correlation).
Shelton additionally reports that a blended tactical allocation
combining S2F with Metcalfe, technical, and sentiment signals does
generate positive out-of-sample returns; we return to this
predictor-combination angle in Section~\ref{sec:open}, since it
relates to open questions rather than to the standalone validity of
S2F.

\paragraph{Metcalfe's Law.}
Metcalfe's Law---the proposition, dating to the 1980s, that network
value scales quadratically with user
count~\citep{metcalfe2013law}---has been widely invoked to justify
Bitcoin valuation models.
\citet{shanaev2019marginal} (an SSRN working paper, but with
rigorous instrumental-variable methodology) use IV on block-level
data for six proof-of-work cryptocurrencies and find that the
resulting relationship between network activity and price is
\emph{spurious}---driven by autocorrelation and endogeneity.  The
positive effects of hashrate and transaction count on price are
non-existent once properly instrumented.
\citet{stylianou2021network} reach a complementary conclusion via a
different route: examining six cryptocurrencies from inception, they
find that network effects are insufficiently consistent to serve as
reliable valuation tools.  Together, these results undermine a common
justification for on-chain signal-based models.

\paragraph{Power law: origin and current claims.}
The observation that Bitcoin's price follows a power law in time
traces to Santostasi's 2014 Reddit post linking Metcalfe scaling to
Bitcoin price, and his 2018 post explicitly identifying the
price-time power law on a log-log
chart.\footnote{G.\ Santostasi, Reddit posts, March 2014
(\url{https://reddit.com/r/Bitcoin/comments/21pujs/}) and
September 2018
(\url{https://reddit.com/r/Bitcoin/comments/9cqi0k/}).
Accessed April 23, 2026.}  This origin in informal venues is
unusual for a quantitative model but is documented above for
attribution.  \citet{burger2019powerlaw}, an independent blog post,
subsequently elaborated the idea into a power-law corridor with
support and resistance bands, a form that now dominates
practitioner-facing visualisations.
\citet{santostasi2026powerlaw} provide a mechanistic derivation
($P(t) \sim t^{5.69}$, $R^2 = 0.961$) by decomposing the exponent
into cubic address growth composed with generalised Metcalfe scaling,
and validate the relationship through scale-invariance tests,
Bayesian stability analysis, and explicit falsifiability criteria.
However, the paper remains a Zenodo preprint.

\paragraph{Power law: what is missing.}
The authors of~\citet{santostasi2026powerlaw} acknowledge that formal
distributional tests per~\citet{clauset2009powerlaw} have not been
applied; no formal residual diagnostics are reported; and the power
law is not compared against alternative trend specifications via
information criteria.  \citet{broido2019scale} apply the full
Clauset methodology to approximately 1{,}000 real-world networks
and find that only 4\% exhibit genuine power-law
structure---underscoring the importance of completing these formal
tests for Bitcoin before drawing strong conclusions in either
direction.

\paragraph{Bubble detection.}
A different structural objective is to identify when prices have
departed from fundamental value, rather than to forecast levels.
\citet{wheatley2019bubbles} combine generalized Metcalfe's Law for
fundamental value with the Log-Periodic Power Law Singularity (LPPLS)
model for bubble detection.  This is a distinct objective from point
forecasting: the model identifies when price has exceeded
fundamental value and provides ex ante crash warnings, but does not
predict price levels at specific horizons.

\subsection{The evaluation methodology crisis}

The most striking finding from our survey is not about Bitcoin but
about the literature itself: the vast majority of papers use
evaluation methods that cannot establish genuine predictive ability.

\citet{mcnally2018predicting}---among the most-cited Bitcoin
prediction papers---evaluates LSTM on a single chronological
train/test split, reporting a headline classification accuracy of
52\% on direction prediction.  This figure is barely above the 50\%
chance level and, evaluated on a single window, lies well within the
statistical noise expected for a binary outcome.
\citet{katsiampa2017volatility}---widely cited for Bitcoin
volatility modelling---selects among GARCH variants using
information criteria (AIC, BIC, Hannan-Quinn) on the full sample,
with no out-of-sample evaluation.
\citet{kose2025deep} test several ML/DL architectures with
macroeconomic drivers but do not include a naive-baseline
comparison; ARIMA serves as their econometric benchmark.
Prior Bayesian Structural Time Series work on
Bitcoin~\citep{poyser2019exploring} focused on in-sample
decomposition without formal out-of-sample evaluation.

The included-studies table (Appendix~\ref{app:studies}) makes this
asymmetry visible: among the 23 studies retained for detailed
discussion, the methodologically rigorous designs (Tier~1) all use
walk-forward, rolling-window, or expanding-window evaluation on a
single contiguous out-of-sample period.  None evaluate across
multiple non-overlapping holdout windows spanning different market
regimes---a distinct and stronger design that detects regime-induced
failure modes that walk-forward, by construction, can blur into a
single error metric.

This echoes a well-known concern in quantitative finance.
\citet{bailey2014pseudo} show that when many parameter combinations
are tested on historical data, the probability of spurious
out-of-sample success grows rapidly.  Much of the Bitcoin prediction
literature---and the Tier-3 studies in our sample in
particular---is susceptible to this problem, as papers typically
select the best-performing model from multiple candidates evaluated
on a single window.

\section{Social media as informal peer review}

Important methodological discourse about Bitcoin models is happening
outside journals, on X/Twitter.  This section is illustrative, not a
systematic content analysis: we present representative examples
identified through targeted Grok searches (Section~2) to document
qualitative patterns in the discourse, not to quantify their
prevalence.  While these discussions lack the structure of formal
peer review, they often raise valid concerns that academia has not
addressed.

\subsection{Valid observations from practitioners}
\label{sec:practitioners}

\paragraph{OLS violations in power-law fitting.}
Quantitative analysts on X have pointed out that OLS regression on
Bitcoin's log-log price data violates all four Gauss-Markov
assumptions simultaneously: the data is non-stationary (declining
volatility over time), autocorrelated (today's price predicts
tomorrow's), right-skewed (bull overshoots exceed bear drawdowns),
and fat-tailed (extreme moves occur far more often than a Gaussian
predicts).\footnote{@TheRealPlanC, X post, March 16, 2026.
\url{https://x.com/TheRealPlanC/status/2033648689328361960}.
Accessed April 21, 2026.}  These are valid statistical
concerns, though they affect inference (confidence intervals,
standard errors) more than point forecasting.

\paragraph{Power-law testing standards.}
Practitioners have referenced \citet{broido2019scale} to argue that
Bitcoin's claimed power law should be subjected to formal statistical
testing rather than accepted on the basis of visual fit in log-log
space.\footnote{@CorySwan, X post, November 15, 2025.
\url{https://x.com/CorySwan/status/1989856322561642709}.
Accessed April 21, 2026.}  This is a legitimate concern that the
academic literature has not addressed.

\paragraph{Backtest overfitting awareness.}
References to \citet{bailey2014pseudo} appear in practitioner
discourse, showing awareness that selecting the best-performing model
from many backtests invalidates out-of-sample
claims.\footnote{@predict\_addict, X post, December 2025.
\url{https://x.com/predict_addict/status/1936107727685595624}.
Accessed April 21, 2026.}

\paragraph{Cross-cryptocurrency spillovers.}
Crypto traders have long operated on the intuition that ``BTC goes
first, alts follow with a delay.''  Analysts track Bitcoin dominance
breakdowns and lag metrics to time altcoin
entries,\footnote{@TechDev\_52, X post, December 29, 2024 (528+
likes, 73{,}000+ views): ``The alt market has peaked after Bitcoin at
macro tops with a consistent lag.''
\url{https://x.com/TechDev_52/status/1873364634117521723}.
@milesdeutscher, X post, February 27, 2024 (440+ likes).
Accessed April 22, 2026.} with some quantifying the lag, though estimates vary by an order
of magnitude---from 6--8 weeks to 6--8
months.\footnote{@CryptoMichNL, X post, April 16, 2026 (191+
likes), estimating 6--8 weeks.
\url{https://x.com/CryptoMichNL/status/2044853894921208267}.
@cas\_abbe, X post, June 10, 2025, estimating 6--8 months.
Accessed April 22, 2026.}  This practitioner knowledge---framed in
terms of ``rotation,'' ``beta,'' and dominance
charts---was formally confirmed by~\citet{guo2024cross}, who document
statistically significant cross-cryptocurrency return predictability
driven by slow information diffusion.  No posts in this spillover
sample reference the academic literature; the insight is purely
empirical and chart-driven, yet it aligns precisely with the formal
econometric finding.

\subsection{Persistence of challenged models and inflated claims}

The opposite dynamic also operates: models debunked in peer-reviewed
work continue to command large audiences on social media, while new
models are promoted with accuracy claims that no peer-reviewed study
supports.  This is a description of a dominant, high-engagement
promotional pattern, not a claim of universal absence of critique:
the critical voices documented in Section~\ref{sec:practitioners}
circulate in the same broader discourse, and some accounts post
both promotional and critical content.  The examples below
illustrate the promotional pattern, not the entirety of
practitioner discourse.

\paragraph{Falsified models.}
The stock-to-flow model's creator continues to promote S2F predictions
of \$250k--\$1M for the 2024--2028 halving cycle, with individual
posts receiving over 500{,}000
views.\footnote{@100trillionUSD, X post, March 8, 2026.
\url{https://x.com/100trillionUSD/status/2030627876698050937}.
Accessed April 22, 2026.}  When original price-level predictions
failed to materialize, the framework's scope shifted from specific
prices to ``halving-period averages'' and from price to
``value'': ``it is value, not price \ldots bitcoin is (extremely)
undervalued.''\footnote{@100trillionUSD, X post, April 7, 2026.
\url{https://x.com/100trillionUSD/status/2041533449115197523}.
Accessed April 22, 2026.}  This redefinition moves the model toward
unfalsifiability---if any deviation is attributed to
price-versus-value rather than model failure, no outcome can
disconfirm it.

These posts are widely shared by crypto news outlets and influencer
accounts with six-figure
followings.\footnote{@BSCNews, X post, March 10, 2026;
@DustyBC, X post, April 22, 2025.  Accessed April 22, 2026.}
No high-engagement post in our sample cited or engaged with the
peer-reviewed work of \citet{shelton2024bitcoin},
which demonstrates S2F's out-of-sample failure using proper
statistical methodology.  The rare
acknowledgements of criticism pivot to other bullish factors
(e.g., institutional adoption) rather than addressing the
statistical evidence.

\paragraph{Metcalfe's Law as live valuation tool.}
Network-value models based on Metcalfe's
Law~\citep{metcalfe2013law} remain actively used for real-time
Bitcoin valuation.  The author of a widely cited Metcalfe valuation
paper treats address-based network value as an undervaluation signal,
claiming 60\%+ rallies within 6--9 months when price dips below
model value.\footnote{@nsquaredvalue (Timothy Peterson), X post,
April 21, 2026 (182+ likes).
\url{https://x.com/nsquaredvalue/status/2046627136207757682}.
Accessed April 22, 2026.}
Other high-engagement accounts apply the quadratic formula directly:
``Bitcoin Price is quadratic with the number of active users \ldots
If that goes to 200MM we get to 1.7MM /
coin.''\footnote{@dotkrueger, X post, September 28, 2025 (638+ likes,
59{,}000+ views).
\url{https://x.com/dotkrueger/status/1972338023783452748}.
Accessed April 22, 2026.}  On-chain analysts
build explicit models from $\text{Price} = A \times
\text{ActiveAddresses}^2 / \text{Supply}$ with calibrated
bands.\footnote{@MrStefirta, X post, April 10, 2026.  Accessed
April 22, 2026.}  None of these posts reference the instrumental
variable analysis of~\citet{shanaev2019marginal}, which found the
Metcalfe relationship to be spurious once endogeneity is addressed.
The model persists because the raw correlation between addresses and
price is visually compelling---precisely the kind of evidence that
formal testing is designed to scrutinize.

\paragraph{Inflated AI/ML accuracy claims.}
A separate promotional ecosystem markets AI-powered Bitcoin prediction
tools with accuracy claims far exceeding anything in the peer-reviewed
literature.  Posts routinely report 95--99\% accuracy for ML models
(LightGBM, ARIMA hybrids, neural networks), often accompanied by
paid signal services or trading
bots.\footnote{Representative examples: @jansen\_ai, X posts,
April 4--13, 2026, reporting 95.8--99.91\% accuracy for
``DE-optimized ARIMA'' and ``SA + LightGBM'' BTC predictions.
Accessed April 22, 2026.}  These figures invariably reflect
in-sample or short-horizon backtested MAPE on price levels---a
metric that is trivially high for any persistent time series, since
tomorrow's price is typically within a few percent of today's.  The
peer-reviewed reality is stark.  For direction prediction,
\citet{arain2025forecast} report out-of-sample directional accuracy
broadly in the low-to-mid 50s (in their tables, peaking around 57\%
for the best predictor and clustering near 50\% for most),
consistent with the headline 52\% of~\citet{mcnally2018predicting}.
For price and return prediction, ML and DL models rarely beat
simpler benchmarks out-of-sample~\citep{puoti2024quantifying,
cakici2024machine}, and even statistically significant edges often
fail to generate economic value after transaction
costs~\citep{arain2025forecast}.  The gap between 99\% backtested
accuracy and a mid-50s out-of-sample directional accuracy captures
the field's central methodological failure in miniature.

\subsection{The gap}

The gap between academic publications and social media discourse runs
in both directions.

On one side, practitioners raise valid concerns that academia has not
formalized: OLS violations in power-law fitting, the need for formal
power-law testing, and awareness of backtest overfitting.  In some
cases---notably cross-cryptocurrency spillovers---practitioner
intuition predates and aligns with formal econometric findings.
These insights could motivate rigorous academic work but remain
informal.

On the other, social media lacks the mechanisms to retire falsified
models.  The S2F model continues to reach hundreds of thousands of
viewers years after peer-reviewed falsification, because social media
has no equivalent of journal retraction or failed replication.
In our sample, promotional content was far more visible than
critical analysis, and we found no serious social media engagement
with the concept of multi-regime evaluation.

The result is a field where valid concerns circulate informally on
one side while falsified models persist with large audiences on the
other, and the academic literature continues to publish papers with
evaluation methodologies that cannot establish genuine predictive
ability.

\section{What is established}

Table~\ref{tab:established} summarizes the current state of
evidence, separating peer-reviewed findings from provisional results
that await independent replication or formal review.

\begin{table}[t]
\centering
\small
\begin{tabular}{p{4.8cm}p{3.8cm}p{3.4cm}}
\toprule
Claim & Evidence & Status \\
\midrule
\multicolumn{3}{l}{\textit{Peer-reviewed findings}} \\
\addlinespace
Naive baseline competitive at short-to-medium horizons (up to 30~d) &
  Puoti~'24, Cakici~'24, Agakishiev~'25 & Supported \\
\addlinespace
Short-term predictability exists (daily) &
  Gradojevic~'23, Berger~'24, Gurgul~'25 & Supported$^*$ \\
\addlinespace
Statistical gains $\neq$ economic gains &
  Arain~'25 & Supported \\
\addlinespace
Stock-to-flow fails out-of-sample &
  Shelton~'24 & Supported \\
\addlinespace
Cross-crypto spillovers exploitable &
  Guo~'24, Magner~'22 & Limited evidence \\
\addlinespace
In-sample predictors fail out-of-sample &
  Yae~'22 & Single study \\
\addlinespace
Metcalfe's Law is spurious or inconsistent &
  Stylianou~'21; Shanaev~'19 (IV)$^\dagger$ & Supported \\
\midrule
\multicolumn{3}{l}{\textit{Provisional findings (preprints and working papers)}} \\
\addlinespace
Naive dominance extends to monthly multi-regime holdout &
  Baquero~'26$^\ddagger$ & Single study \\
\addlinespace
Power law describes long-run trend (empirical fit) &
  Santostasi~'26$^\ddagger$ & Empirical fit only$^\S$ \\
\bottomrule
\end{tabular}
\caption{Summary of evidence.  Upper panel: findings from
  peer-reviewed studies (plus one working paper, Shanaev, used in
  conjunction with peer-reviewed Stylianou).  Lower panel:
  provisional findings from preprints and working papers that await
  peer review.  Status labels: \emph{Supported} = multiple
  independent studies in agreement; \emph{Single study} or
  \emph{Limited evidence} = one or two studies, no independent
  replication; \emph{Empirical fit only} = strong
  visual/parametric fit but missing formal distributional and
  comparative tests.
  $^*$Frequency-specific: daily predictability does not extend to
  hourly~\citep{gradojevic2023forecasting} or, on present evidence,
  to monthly horizons.
  $^\dagger$SSRN working paper with rigorous IV methodology.
  $^\ddagger$Preprint, currently under peer review.
  $^\S$Scale-invariance tests and Bayesian stability analysis
  performed; formal distributional tests
  per~\citet{clauset2009powerlaw}, residual diagnostics, and
  alternative-model comparison still lacking.}
\label{tab:established}
\end{table}

The asymmetry is notable: the field has many more failed or
challenged claims than validated positive results.  The strongest positive
result---short-term microstructure predictability---is
frequency-specific and may not survive transaction
costs~\citep{arain2025forecast}.  At the horizons most investors
care about (months to years), the best-supported baseline remains
today's price: no peer-reviewed alternative has yet shown robust
superiority.  The clearest candidate alternative at multi-year
horizons is the power-law trend, which fits the historical price
series tightly but has not yet been subjected to the formal
distributional tests that the field's own methodological standards
require~\citep{clauset2009powerlaw}---it remains an empirical
regularity awaiting peer-reviewed validation, not a confirmed
predictive model.

\section{Open questions and research directions}
\label{sec:open}

Several genuinely open questions emerge from this survey, each
representing a viable research direction.

\paragraph{Can any model beat naive at 1--6 months across multiple
regimes?}
This is the central open question.  Only one study has tested this
with multi-regime holdout~\citep{baquero2026naive}, finding that no
model survives.  Independent replications are needed with different
model families (deep learning, gradient boosting, reinforcement
learning), data sources (macroeconomic indicators, sentiment, order
flow), and evaluation protocols.

\paragraph{Non-linear models at monthly horizons.}
\citet{baquero2026naive} tested linear corrections and BSTS.
Non-linear models (neural networks, gradient boosting) might capture
interactions that linear models miss.  However, the finding that
Bayesian model averaging also selects zero suggests a fundamental
bias-variance issue: at monthly horizons, the variance cost of any
non-zero correction exceeds the bias reduction.  This hypothesis
needs empirical testing.

\paragraph{External data sources.}
Macroeconomic variables (VIX, interest rates, dollar index), natural
language sentiment, and exchange-level order flow remain untested
with proper multi-regime evaluation.  \citet{kose2025deep} test
macro drivers but do not benchmark against a naive baseline.  The
finding of~\citet{yae2022outofsample} that in-sample predictors
fail out-of-sample, and of~\citet{baquero2026naive} that on-chain
signal selection is regime-specific, suggests that external signals
are likely subject to the same non-stationarity problem and should
be evaluated under the same multi-regime protocol.

\paragraph{Declining volatility and changing predictability.}
Bitcoin's volatility has declined substantially since its early
years~\citep{baur2021volatility}, and~\citet{baquero2026naive}
document a shortening crossover horizon at which the power-law trend
becomes more informative than price persistence.  If this trend
continues, deterministic trend models may become useful at shorter
horizons in future regimes---a testable prediction.

\paragraph{Predictor combination.}
Individual structural and on-chain predictors have repeatedly
failed standalone out-of-sample tests, yet~\citet{shelton2024bitcoin}
reports that a blended tactical allocation combining several weak
predictors (S2F, Metcalfe, technical, and sentiment signals)
generates positive out-of-sample returns.  This raises a distinct
question: even if no individual signal beats the naive baseline,
can a principled combination of weak signals do so consistently
across regimes?  The current evidence is suggestive but rests on a
single study; testing combination strategies under multi-regime
holdout, with explicit guards against the
backtest-overfitting risk identified by~\citet{bailey2014pseudo},
is an open direction.

\paragraph{Power-law validation.}
The scale-invariance tests and Bayesian stability analysis
of~\citet{santostasi2026powerlaw} are valuable, but the authors
acknowledge that formal
distributional tests per \citet{clauset2009powerlaw}---maximum
likelihood fitting, KS goodness-of-fit, and likelihood ratios
against alternatives---have not been applied to the temporal series.
Equally needed are formal residual diagnostics (stationarity,
autocorrelation, heteroscedasticity) and comparison against
alternative trend specifications (logarithmic, piecewise linear,
polynomial) using information criteria and out-of-sample forecasting.
\citet{broido2019scale} applied distributional tests to
$\sim$1{,}000 networks and found that only 4\% exhibited genuine
power-law structure.  Completing these tests for Bitcoin would resolve
whether the power law is a true scaling law or merely a
convenient approximation.

\section{Methodological recommendations}

Based on our survey and established principles of financial
econometrics~\citep{campbell1997econometrics}, we propose the
following standards for future Bitcoin price prediction research.

\subsection*{Mandatory requirements}

Any paper claiming predictive ability for Bitcoin should satisfy all
of the following:

\begin{enumerate}
  \item \textbf{Walk-forward or rolling-window evaluation.}  Single
    train/test splits are insufficient for non-stationary data.  The
    model must be evaluated at multiple points in time, with training
    data expanding or rolling forward.

  \item \textbf{Multi-regime holdout windows.}  Define at least three
    non-overlapping holdout periods spanning different market
    conditions (bull, bear, sideways).  A model that works in one
    regime but fails in another is not a reliable model.

  \item \textbf{Naive baseline comparison.}  Every model must be
    compared to ``today's price'' (for price-level prediction) or
    ``zero return'' (for return prediction).  This is the most basic
    test and is frequently omitted.

  \item \textbf{Include zero in hyperparameter grids.}  If a model
    has a correction-strength parameter, zero (no correction) must be
    a candidate.  This is equivalent to including the null hypothesis
    in the model selection~\citep{baquero2026naive}.

  \item \textbf{Statistical significance testing.}  Forecast
    differences must be tested for significance using the
    Diebold-Mariano test~\citep{diebold1995comparing} for pairwise
    comparisons, and the Model Confidence Set
    of~\citet{hansen2011mcs} when multiple competing models are
    evaluated jointly; not merely reported as point differences.
\end{enumerate}

\subsection*{Recommended practices}

\begin{enumerate}[resume]
  \item \textbf{Economic significance testing.}  Even statistically
    significant improvements may be economically irrelevant after
    transaction costs~\citep{arain2025forecast}.  Report whether
    forecast gains translate to trading profits.

  \item \textbf{Report forecast task and horizon explicitly.}  State
    both the task (price-level forecasting, return forecasting, or
    direction classification; cf.\ Table~\ref{tab:tasks}) and the
    horizon prominently.  Daily predictability does not imply
    monthly predictability, and a model that beats the naive
    baseline for one task may fail for another; do not generalize
    beyond what was tested.

  \item \textbf{Guard against backtest overfitting.}  When many model
    configurations are evaluated, the probability of spurious
    out-of-sample ``success'' increases
    rapidly~\citep{bailey2014pseudo}.  Report the number of
    configurations tested and adjust claims accordingly.

  \item \textbf{Report negative results.}  The field suffers from
    publication bias: models that fail to beat naive are less likely
    to be published.  Negative results are informative and should be
    publishable.  We note that acting on this recommendation
    requires editorial and reviewer norms to shift; authors who
    submit well-executed null results should be supported, not
    penalised, by referees.

  \item \textbf{Separate discovery from evaluation, and guard
    against look-ahead leakage.}  If hyperparameters are tuned on
    data, that data cannot be part of the holdout evaluation.  A
    common failure mode in Bitcoin ML pipelines is using
    full-sample statistics (means, standard deviations, technical
    indicator normalisations) at training time, which silently
    leaks information from the test period.  Compute all
    preprocessing statistics from the training window only.

  \item \textbf{Release code and data.}  Publish the analysis code,
    preprocessing pipeline, and the exact data splits used.  This is
    the only reliable way to permit independent replication on a
    non-stationary asset where historical data is continually
    revised and re-aggregated by data providers.
\end{enumerate}

\section{Conclusion}

The peer-reviewed landscape of Bitcoin price prediction is dominated
by volume rather than validated results.  Hundreds of papers apply
machine learning, deep learning, and econometric models, but the vast
majority use evaluation methods that cannot establish genuine
predictive ability: single splits, in-sample metrics, no naive
baseline.

When evaluated rigorously---with walk-forward protocols, multi-regime
holdout windows, and naive baseline comparisons---the best current
evidence is sobering.  At 1--6 month horizons, no peer-reviewed
study has demonstrated a model that reliably beats the naive forecast
(today's price) across multiple market regimes.  Daily
predictability is real but does not extend to hourly or monthly
horizons, and may not survive transaction costs.  The stock-to-flow
model has failed formal out-of-sample testing, and Metcalfe's Law
valuations have been challenged as spurious.  The Bitcoin price
power law, while empirically compelling, has not been subjected to
formal distributional tests and remains the clearest candidate
alternative to the naive baseline at multi-year horizons---a
candidate awaiting peer-reviewed validation, not a confirmed
predictive model.

Meanwhile, the gap between academic publications and social media
discourse runs in both directions.  Practitioners raise valid
methodological concerns---OLS violations, backtest overfitting,
spurious regressions---that the academic literature has not
formalized, and in some cases (cross-cryptocurrency spillovers)
practitioner intuition predates formal econometric confirmation.
Yet social media also lacks mechanisms to retire falsified models:
stock-to-flow and Metcalfe-based valuations continue to reach
hundreds of thousands of viewers years after peer-reviewed
challenges, while AI prediction services claim 95--99\% accuracy
that no Tier~1 study supports.

The field's primary need is not more models but better evaluation.
Walk-forward protocols, multi-regime holdout windows, naive
baseline comparisons, zero-correction inclusion in hyperparameter
grids, and Diebold-Mariano (or Model Confidence Set) significance
testing should be minimum standards.  The most important open
question is not which model is best, but whether \emph{any} model,
using any data source, can reliably beat the naive baseline at
short-to-medium horizons across multiple market regimes.

\bibliographystyle{plainnat}
\bibliography{landscape_references}

\appendix

\newpage

\section{Included studies}
\label{app:studies}

Table~\ref{tab:studies} lists the Bitcoin prediction studies
discussed in this survey, with key characteristics.  Studies were
identified through targeted searches on Google Scholar and Web of
Science (query: ``Bitcoin price prediction/forecasting'',
2017--2026, searched April 2026), supplemented by forward/backward
citation tracing and domain expertise.  Approximately 120 initial results were screened by title and
abstract; full texts were retrieved for approximately 40 candidates.
We retained 23 studies for detailed discussion: those with explicit
out-of-sample evaluation, widely cited studies included to illustrate
methodological weaknesses, and structural/valuation models central to
the social media discourse.  Foundational references (e.g.,
Clauset et al., Meese \& Rogoff, Bailey et al.) are cited for
methodology but not listed in the study table.
Tier assignments follow the criteria in Section~2.

\begin{table}[h!]
\centering
\footnotesize
\setlength{\tabcolsep}{2.5pt}
\begin{tabular}{p{2.5cm}p{1.4cm}p{1.3cm}p{1.3cm}p{1.6cm}p{2.0cm}p{1.3cm}c}
\toprule
Study & Task & Target & Horizon & Sample & Evaluation & Baseline & Tier \\
\midrule
Puoti~'24 & PL & Price & 1--30\,d & 2020--23 & Walk-fwd & Naive & 1 \\
Baquero~'26$^*$ & PL & Price & 1--6\,mo & 2016--26 & Multi-reg. & Naive & 1 \\
Arain~'25 & RF & Returns & 1\,d & 2021--24 & Expanding & Naive & 1 \\
Yae~'22 & RF & Returns & 1\,d & 2017--20 & Expanding & Naive & 1 \\
Cakici~'24 & RF & Returns & X-sec. & 2017--23 & OOS split & OLS & 1 \\
Agakishiev~'25 & RF & Portf. & Daily & 2015--22 & OOS split & Simple & 1 \\
Gradojevic~'23 & RF & Returns & 1\,h, 1\,d & 2015--19 & Expanding & RW & 1 \\
Berger~'24 & RF & Returns & 1--10\,d & 2013--21 & Rolling & Naive & 1 \\
Kim~'25 & DC & Direct. & 1\,d & 2015--22 & OOS split & --- & 2 \\
Wei~'23 & RF & Returns & 1\,d & 2014--19 & Rolling & Best lin. & 1$^\dagger$ \\
Gurgul~'25 & DC & Direct. & 1\,d & 2011--23 & Rolling CV & B\&H & 1$^\dagger$ \\
Guo~'24 & RF & Returns & X-sec. & 2019--23 & OOS split & --- & 1 \\
Shelton~'24 & RF & Returns & Mo. & 2014--24 & Expanding & OOS $R^2$ & 1 \\
Shanaev~'19$^*$ & SV & Price & --- & Block & IV regr. & OLS & 1 \\
Santostasi~'26$^*$ & SV & Price & --- & 2010--26 & In-samp.+scal. & --- & 3 \\
Wheatley~'19 & BD & Bubbles & --- & 2010--18 & Regime det. & No-bub. & 2 \\
McNally~'18 & DC & Direct. & 1\,d & 2013--16 & Single split & ARIMA & 2 \\
Katsiampa~'17 & SV & Volat. & --- & 2010--16 & AIC/BIC & --- & 3 \\
K\"{o}se~'25 & PL & Price & 1\,d & 2012--24 & Single split & ARIMA & 2 \\
Poyser~'19 & SV & Decomp. & --- & 2010--17 & In-sample & --- & 3 \\
Magner~'22 & RF & Returns & 1\,d & 2018--22 & OOS split & RW & 1 \\
De Nicola~'21 & RF & Returns & 1--4\,h & 2013--20 & In-sample & --- & 3 \\
Stylianou~'21 & SV & Network & --- & Inception & Empirical & --- & 3 \\
\bottomrule
\end{tabular}
\caption{Characteristics of included studies.
  Task codes: PL = price-level forecasting, RF = return forecasting,
  DC = direction classification, SV = structural valuation,
  BD = bubble detection (cf.\ Table~\ref{tab:tasks}).
  $^*$Preprint or working paper.
  $^\dagger$No naive/random-walk baseline; tier reflects other
  methodological strengths.
  ``X-sec.''\ = cross-sectional.
  ``RW'' = random walk.
  ``B\&H'' = buy-and-hold.
  ``---'' = not applicable or not reported.}
\label{tab:studies}
\end{table}

\newpage

\section{AI use disclosure}
\label{app:ai}

AI tools were used to assist with literature search, drafting, and
consistency checks throughout this work.

\paragraph{Literature search and verification.}
Claude (Anthropic, models in the Claude Opus~4 family, including
\texttt{claude-opus-4-6} and \texttt{claude-opus-4-7}) was used
interactively to assist with: searching for and retrieving academic
papers, verifying bibliographic metadata (authors, volumes, DOIs,
publication status), drafting reference summaries, and identifying
gaps in the survey coverage.  Verification against the source PDFs
of factual claims (numerical values, direct quotations, study
methodology) was performed by the author with Claude assistance.
Social media discourse was identified using Grok (X's AI
assistant) for targeted post searches, with results verified by
the author.

\paragraph{Writing and revision.}
Claude assisted with drafting text, restructuring sections in
response to reviewer feedback, checking internal consistency
(cross-references, citation completeness, terminology), and
generating the study characteristics table in
Appendix~\ref{app:studies}.  Additionally, GPT (OpenAI, GPT-5.4)
was used to produce independent reviewer-style feedback on
intermediate drafts; this feedback was assessed and selectively
incorporated by the author.

\paragraph{Scientific decisions.}
All scientific judgements---including the tier classification system,
the selection and interpretation of included studies, the framing of
the social media analysis, and the methodological
recommendations---were made by the human author.  AI tools were used
as assistants and do not meet the criteria for authorship.

\paragraph{Responsibility.}
In accordance with ICMJE guidelines, the human author bears full
responsibility for the integrity of the survey, the correctness of
the reported findings, and the content of the manuscript.  

\end{document}